\documentclass{elsart}
\usepackage[latin1]{inputenc}
\usepackage{epsfig}
\usepackage{amsfonts}
\usepackage{amssymb}
\newcommand{\beq}{\begin{equation}}
\newcommand{\eeq}{\end{equation}}
\newcommand{\la}{\langle}
\newcommand{\ra}{\rangle}

\begin{document}

\begin{frontmatter}

\title{Positive heat capacity in the microcanonical ensemble}

\author{M\'{a}rio J. de Oliveira}

\address{Instituto de F\'{\i}sica,
Universidade de S\~{a}o Paulo, \\
Rua do Mat\~ao, 1371,
05508-090 S\~{a}o Paulo, S\~{a}o Paulo, Brazil}

\begin{abstract}

The positivity of the heat capacity is the hallmark
of thermal stability of systems in thermodynamic equilibrium.
We show that this property remains valid 
for systems with negative derivative of energy with respect
to temperature, as happens to some system described by
the microcanonical ensemble. The demonstration rests on considering
a trajectory on the Gibbs equilibrium surface,
and its projection on the entropy-energy plane.
The Gibbs equilibrium surface has the convexity property, but the projection
might lack this property, leading to a negative derivative
of energy with respect to temperature.

\end{abstract}


\end{frontmatter}

\section{Introduction}

Heat capacity is the
ratio between the heat introduced in a system and the increase in
its temperature, $C=dQ/dT$. The infinitesimal heat $dQ$ is not
an exact differential but, according to Clausius, there exists an integrating
factor, the inverse of the temperature, that makes $dQ$ an exact differential.
The resulting exact differential allows the definition of
entropy, $dS=dQ/T$, and the heat capacity becomes $C=T(dS/dT)$.
The relation $dQ=TdS$ is valid as long as the system
is in equilibrium. In out of equilibrium, although one may
still assign an entropy to the system, such a
relation does not hold because temperature cannot be 
unambiguously assigned to a non-equilibrium system.
Nevertheless,  the ratio $dQ/T$ can be determined
if $T$ is understood as the temperature of the environment
with which the system is in contact. In this case,
according to Clausius, the quantity $dQ/T$ is not equal to $dS$ but
is smaller due to the generation of entropy inside the system. 
Defining the heat flux $\Phi_q$ as the heat introduced into
the system per unit time, the time variation of the entropy of
the system is given by the Clausius inequality $dS/dt \geq \Phi_q/T$, 
which is a statement of the second law of thermodynamics.

Defining the entropy flux, that is, the entropy flow into
the system per unit time by $\Phi=\Phi_q/T$,
where $T$ is again the temperature of the environment,
the Clausius inequality can be written as \cite{tome2015}
\beq
\frac{dS}{dt} = \Pi + \Phi,
\label{4}
\eeq
where $\Pi$ is the rate of entropy production,
and the statement of the second law becomes $\Pi\ge0$.
The main consequence of $\Pi\ge0$ combined with Eq. (\ref{4}) 
is the inequality concerning the heat capacity, $C\geq0$.
This fundamental inequality is equally the hallmark of the
thermal stability \cite{landau1958}.
More precisely, it
is a consequence of the convexity of the thermodynamic
potentials \cite{oliveira2015}, which in turn is a direct 
result coming from the second law expressed by $\Pi\ge0$.
A stable system is thus characterized by 
a nonnegative heat capacity. 
According to Landau and Lifshitz, equilibrium
states that do not fulfill this condition are in fact 
unstable and cannot exist in Nature \cite{landau1958}.

Our aim here is to emphasize
the positivity of the heat capacity in situations in which
$\partial U/\partial T$, the derivative of the energy $U$ with respect
to the temperature $T$, is negative, which at first-sight seems to 
yield a negative heat capacity. 
The main example of these situation is a small system described by
the microcanonical ensemble. When $T$ is plotted against $U$, it is found
that there is an interval in $U$ in which $\partial U/\partial T$
is negative
\cite{hertel1971,gross1996,ispolatov2001,gross2001,dauxois2002,thirring2003,%
gross2005,behringer2006,dunkel2006,martin2007,fiore2009,carignano2010,%
schnabel2011,troster2012}.

\section{Convexity}

The positivity of the heat capacity is a direct consequence of the
convexity of the Gibbs surface, which is the surface of equilibrium states in
the space of the thermodynamic extensible variables. 
The convexity property can be derived from the inequality $\Pi\ge0$
and Eq. (\ref{4}) as follows. We start by
considering the variation of the energy $U$ of a
system. The increase of energy per unit time is due to the 
heat flux $\Phi_q$ plus the work done on the system per unit time,
or power, $\Phi_w$,
\beq
\frac{dU}{dt} = \Phi_q + \Phi_w.
\label{7}
\eeq
Generically, the power is written as a field variable $y$
multiplied by $dX/dt$, the time variation of an extensible
variable $X$, that is, $\Phi_w = y dX/dt$.

\begin{figure}
\epsfig{file=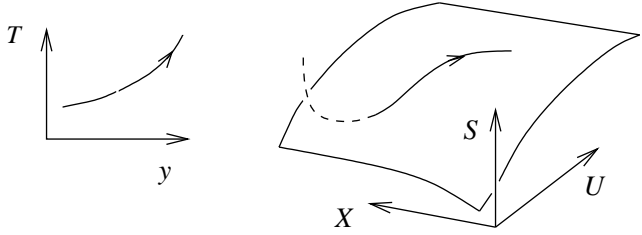,width=8.5cm}
\caption{A path in the $(T,y)$ space and the corresponding 
trajectory in the thermodynamic space $(S,U,X)$. When
$T$ and $y$ vary very slowly, the trajectory in the thermodynamic
space approaches and remains on the Gibbs equilibrium surface
defined by (\ref{10}).}
\label{traj}
\end{figure}

The replacement of $\Phi=\Phi_q/T$ into (\ref{4}) gives
\beq
T\frac{dS}{dt} = T\Pi + \Phi_q,
\label{6}
\eeq
which can be written as
\beq
\frac{dU}{dt} - T\frac{dS}{dt} - y\frac{dX}{dt} = - T\Pi.
\label{8}
\eeq
Notice that $T$ and $y$ refer to the temperature and field of
the environment and not of the system.
Let us suppose that the field variables $T$ and $y$ vary in time
very slowly causing small variations in energy, entropy and $X$.
The point representing the system in the thermodynamic space
$(U,S,X)$ will describe a trajectory that approaches the
equilibrium surface described by
\beq
dU - TdS - ydX =0,
\label{10}
\eeq
as illustrated in Fig. \ref{traj}.
This so happens because the rate of entropy production $\Pi$ becomes
negligible when compared with the time variation of $U$, $S$, and $X$.
That is, the rate of entropy production is of the order greater than
that of the time variation of $U$, $S$, and $X$.
The right-hand side of (\ref{8})
may thus be set to zero resulting in Eq. (\ref{10}), which
tells us that, in the equilibrium regime, $T$ and $y$ become 
the tangents to the Gibbs surface, that is,
\beq
T = \left(\frac{\partial U}{\partial S}\right)_X, \qquad\qquad
y = \left(\frac{\partial U}{\partial X}\right)_S,
\eeq
and we may recognize $T$ and $y$ as being the temperature 
and field of the system in equilibrium in addition to being
the temperature and field of the environment.

To show that the Gibbs surface has the property of convexity
we proceed as follows.
Let the temperature and field at the point $(U_0,S_0,X_0)$ of 
the Gibbs surface be $T_0$ and $y_0$, respectively. 
Suppose that the system evolves with the temperature and field
being kept constant at the values $T_0$ and $y_0$. 
Starting from a state $(U_1,S_1,X_1)$ at time $t=0$, the system evolves in time
and eventually reaches the state $(U_0,S_0,X_0)$. Integrating
Eq. (\ref{8}) in time from zero to infinity, one finds
\beq
(U_0 - U_1) - T_0(S_0-S_1) - y_0(X_0-X_1) = - T_0\int_0^\infty \Pi dt.
\label{11}
\eeq
Considering that $\Pi\geq0$, the right-hand side is smaller
or equal to zero and one reaches the result
\beq
(U_1 - U_0) - T_0(S_1-S_0) - y_0(X_1-X_0) \geq 0.
\label{11b}
\eeq
Since the initial state $(U_1,S_1,X_1)$ is arbitrary, we may choose it as a point on the
Gibbs surface. With this choice, relation (\ref{11b}) becomes the 
condition for convexity of the Gibbs surface.

From the convexity property of the Gibbs surface, we reach the conditions
of stability \cite{oliveira2015}
\beq
C_X = T\left(\frac{\partial S}{\partial T}\right)_X \geq 0,
\qquad\qquad 
C_y = T\left(\frac{\partial S}{\partial T}\right)_y \geq 0,
\eeq 
where $C_X$ and $C_y$ are the heat capacity at constant $X$
and constant $y$, respectively. 
The convexity property of the Gibbs surface implies that
the thermodynamic potentials $F(T,X)$ and $G(T,y)$, obtained
by successive Legendre transformation from $U(S,X)$, are
concave functions of $T$, implying the two conditions above.
The first condition refers
to stability against thermal perturbation for which 
the extensible variable $X$ remains invariant and the
second when the field variable $y$ is kept constant.
No matter which variable is held constant, field or extensible,
the heat capacity is nonnegative.

It should be remarked that the heat capacity is always 
$T\partial S/\partial T$. It may be identified as $\partial U/\partial T$
only in the case of the absence of macroscopic work, which occurs
when all extensible variables are kept constant. 
In the case of just one extensible variable $X$ in addition to
the energy, it follows from (\ref{10}) that
\beq
T\left(\frac{\partial S}{\partial T}\right)_X
= \left(\frac{\partial U}{\partial T}\right)_X.
\label{29}
\eeq 

\begin{figure*}
\epsfig{file=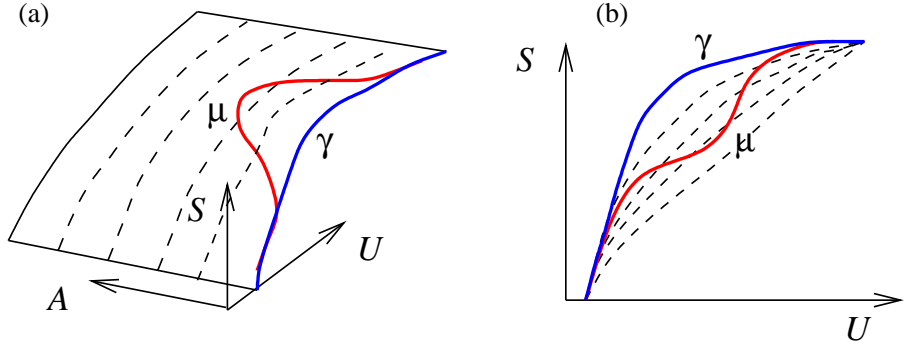,width=12cm}
\caption{(a) Entropy $S$ as a function of $(U,A)$. The $\mu$ trajectory on
the surface represents a microcanonical path, whereas the $\gamma$
trajectory represents a canonical path. The dashed lines
represents curves with constant $A$. (b) Projection on the $(S,U)$
plane.    
}
\label{gibbsurf}
\end{figure*}

\begin{figure*}
\epsfig{file=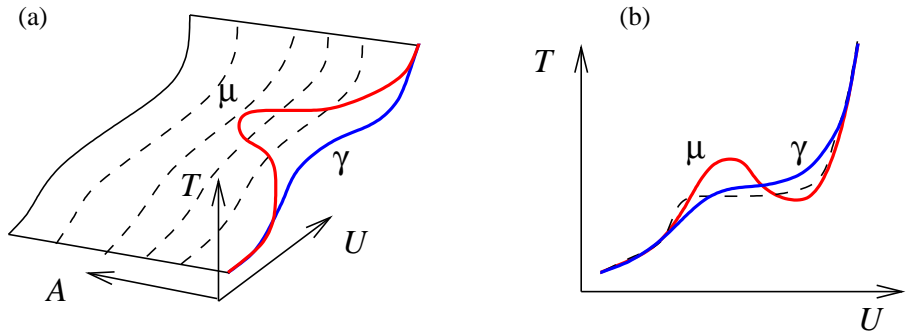,width=12cm}
\caption{(a) Temperature $T$ as a function of $(U,A)$. The $\mu$ trajectory on
the surface represents a microcanonical path, whereas the
$\gamma$ trajectory represents a canonical path. The dashed lines
represents curves with constant $A$. (b) Projection on the $(T,U)$
plane.    
}
\label{tempsurf}
\end{figure*}

\section{Surface of tension}

In the interval of energies where $\partial U/\partial T$
is negative, such as that given by microcanonical calculations,
there is a loop in the curve of temperature versus energy.
In the thermodynamic limit the loop gives away and is
replaced by a tie line, a straight line segment along which the
temperature is constant, indicating the coexistence
of thermodynamic phases. It is natural to presume that
the system in this situation is not homogeneous, 
exhibiting coexisting heterogeneous regions 
with an interface of tension between them \cite{troster2012,zhou2019}.
In accordance with this point of view, the increase in energy of
a system is equal to the heat introduced plus the
the work performed by the surface tension.
In differential form \cite{rowlinson1982},
\beq
dU = TdS + \sigma dA,
\label{12}
\eeq
where $\sigma$ is the surface tension and $A$ is the area
of the interface. Eq. (\ref{12}) describes the Gibbs
equilibrium surface shown in Fig. \ref{gibbsurf}a, which
holds the property of convexity.

Within the microcanonical ensemble, the energy $U$ and
other extensible variables are kept constant, and, according
to Eq. (\ref{29}), the heat capacity would coincide
with the variation of the energy with temperature. However,
the system described by the microcanonical ensemble
might develop internal structures, characterized by extensible variables
that are not or could not be kept constant. An example of
this structure is the interface between two coexisting
thermodynamic phases, characterized by its area.
Therefore, the variation of energy with temperature 
may not coincide with the heat capacity because the area
of the interface, which is an extensible variable, is not
constant and we could not use Eq. (\ref{29}).

In the microcanonical ensemble, the entropy $S$
is determined from partition function $\Omega$ through
the Boltzmann formula $S=k_B \ln\Omega$, and
the area $A$ of the interface could also be determined.
As one increases the energy $U$ from small values,
$S$ and $A$ will vary, and a trajectory is traced
on the Gibbs surface as shown in Fig. \ref{gibbsurf}a,
which we call a trajectory $\mu$. The projection of
the trajectory $\mu$ on the plane $(S,U)$ may lack
the convexity property as seen in Fig. \ref{gibbsurf}b.

From the entropy $S$, and in accordance with Eq. (\ref{12}),
the temperature is determined by
\beq
T=\left(\frac{\partial U}{\partial S}\right)_A,
\label{27}
\eeq
and, knowing $U$ and $A$, we may
draw the trajectory $\mu$ shown in Fig. \ref{tempsurf}a.
The projection of the trajectory $\mu$ in the plane $(T,U)$
may not be monotonic as seen in Fig. \ref{tempsurf}b. 
This explain the negative value of $\partial U/\partial T$ observed in the
microcanonical calculations, but this quantity is not the
heat capacity. In actual microcanonical numerical simulations,
the temperature is not determined by Eq. (\ref{27}),
which would be unpractical,
but by alternative schemes which may or may not coincide with 
formula (\ref{27}). For instance, in simulations of classical
systems of interacting particles it is usual to determine
the temperature by assuming that it is proportional to the 
average of the kinetic energy.

Along the microcanonical trajectory $\mu$, the heat capacity
$C_\mu = T(\partial S/\partial T)_\mu$
is not equal to $(\partial U/\partial T)_\mu$, in general.
Indeed, from Eq. (\ref{12}),
\beq
\left(\frac{\partial U}{\partial T}\right)_\mu
= C_\mu+\sigma\left(\frac{\partial A}{\partial T}\right)_\mu,
\label{19}
\eeq
and $(\partial U/\partial T)_\mu$ is {\it not} the heat capacity 
and may be negative if $(\partial A/\partial T)_\mu$ is negative.
If we define $\lambda=(\partial A/\partial U)_\mu$, which measures the change of the area
with the energy along the trajectory $\mu$, it follows from (\ref{19}) that
\beq
\left(\frac{\partial T}{\partial U}\right)_\mu = \frac{1-\sigma\lambda}{C_\mu}.
\label{20}
\eeq
As one increases the energy starting from small values,
the area $A$ of the interface begins to increase from zero,
reaches a maximum, and then decreases and vanishes again.
At the beginning, $\lambda$ is positive, then vanishes,
and then becomes negative. In the interval where $\lambda$ is positive,
if it is large enough, the quantity $(\partial T/\partial U)_\mu$,
which is the slope of the microcanonical curve of Fig.
\ref{tempsurf}b, will be negative.

In the canonical ensemble, the temperature $T$, which is a parameter, 
and the extensible variables other than energy are kept constant. 
As one varies the parameter $T$, a trajectory is traced
on the surfaces shown in Figs. \ref{gibbsurf}a and \ref{tempsurf}a,
which we call a trajectory $\gamma$.
The entropy is determined by the Gibbs expression
\beq
S = - k_B \int P\ln P dxdp,
\eeq
where $P$ is the probability density defined on the phase space $(x,p)$
and the energy $U$ is the average of the energy function.

The following relation
exists between the entropy and the energy, $S=(U/T)+k_B\ln Z$,
where $Z$ is the canonical partition function. From this
relation we get
\beq
\left(\frac{\partial U}{\partial T}\right)_\gamma
= T\left(\frac{\partial S}{\partial T}\right)_\gamma,
\label{19a}
\eeq
and we may conclude by comparison with the relation analogous
to (\ref{19}) that $(\partial A/\partial T)_\gamma=0$ in the
canonical ensemble, justifying the constance of $A$ in the
trajectory $\gamma$, shown in Figs. \ref{gibbsurf}a and \ref{tempsurf}a.

The right-hand side of Eq. (\ref{19a}) is the heat capacity
$C_\gamma$ along the canonical trajectory and in this case 
\beq
C_\gamma=\left(\frac{\partial U}{\partial T}\right)_\gamma,
\eeq
that is, the heat capacity is identified with the slope
of $U$ versus $T$. Within the canonical ensemble, 
$(\partial U/\partial T)_\gamma$ is proportional to the variance
of the energy function and $C_\gamma$ is a nonnegative quantity as demanded
by the property of convexity of the Gibbs surface.

\begin{figure}
\epsfig{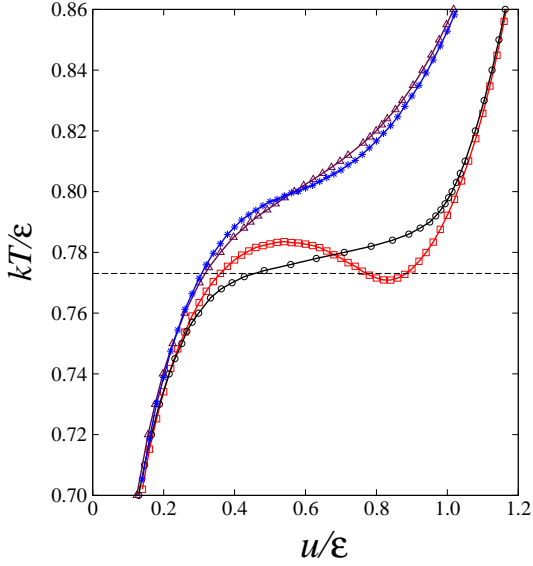}
\caption{Temperature $T$ as a function of the energy per 
site $u$ for the seven-state Potts model on a square lattice
with $N=400$ sites, obtained from the canonical
(circles and triangles) and microcanonical (squares and stars)
ensembles.  The horizontal dashed line represents the
temperature of coexistence in the thermodynamic limit,
$T_0=0.773058$. The two curves (circles and squares) were
obtained with periodic boundary conditions. The other two
curves (triangles and stars) were obtained with fixed
boundary conditions.}
\label{potts}
\end{figure}

\section{Potts model}

The Potts model \cite{wu1982} is defined on a regular lattice in which each
site can be in one of $q$ states. The interaction between two nearest
neighbor sites is $\varepsilon>0$ if the sites are in different states
and zero if they are in the same state. In two dimensions it is known that
a phase transition takes place at the temperature
$k_B T/\varepsilon=1/\ln(1+\sqrt{q})$, which is discontinuous if $q>4$.
This is the case of the seven-state model on
a square lattice, which we focus here.

In the canonical simulations, in which $T$ is a fixed parameter,
we have employed the standard Metropolis algorithm and determined
the energy $U$ as the average of the energy function.
In the microcanonical simulations, we used transition rules
that keep the energy function strictly constant. At each time step
of the simulation, two sites of the lattice are chose at random
and trial states chosen at random are assigned to the sites. 
If the energy remains the same, the trial states become the 
new states of the two sites. The temperature is not obtained
by formula (\ref{27}), which would be unpractical, 
but by a procedure that assumes a local
canonical distribution as follows \cite{shida2003,fiore2006}.
Let us consider a configuration
of the lattice and look for all sites whose neighboring sites
are in the same state. Among the sites of this type, we distinguish
those which are in the same state as its neighbors, and those
which are in a state distinct form its neighbors. We denonte
by $n_0$ the number of site of the former type and by $n_1$
that of the later type. If we use the canonical ensemble 
it is straightforward to show that the ratio of their averages is
is given by
\beq
\frac{\la n_0\ra}{\la n_1\ra} = e^{-4\varepsilon/k_B T}.
\label{31}
\eeq
This formula is then used in microcanonical ensemble  
to calculate the temperature by 
considering that the averages are determined from the
microcanonical simulations. 

Fig. \ref{potts} shows the temperature versus the energy
for the standard seven-state Potts model on a finite square lattice,
which we have obtained by Monte Carlo simulations by using
the microcanonical and canonical ensembles, and two types
of boundary conditions. One of them is the periodic
boundary conditions. In the other type, which
we call fixed boundary conditions, all sites at the boundary
remains permanently in one of the seven states. 
When we use the microcanonical ensemble
and periodic boundary conditions, there is an interval in the energy
for which $\partial U/\partial T$ is indeed negative,
as can be seen in Fig. \ref{potts}. Notice that, 
this does not happen for the microcanonical ensemble and
fixed boundary conditions, and for the canonical ensemble
for both conditions. In all these three cases, 
the temperature is a monotonic increasing function of the energy,
as seen in Fig. \ref{potts}.

The loop observed in the curve of Fig. \ref{potts} 
disappears in the thermodynamic limit giving raise to
a tie line. Assuming that the area of the interface
scales like $N^\alpha$, with $\alpha<1$, the quantity
$\lambda$ in Eq. (\ref{20}) scales like $N^{\alpha-1}$
and vanishes in the thermodynamic limit, and 
$(\partial T/\partial u)_\mu$ approaches $1/c_\mu$
where $c_\mu=C_\mu/N$ is the specific heat. 
In fact, for values of $u$ within the tie line,
both quantities approach the zero value. 
The deviation of $T$ from $T_0$ also scales like $N^{\alpha-1}$.
The exponent $\alpha$ is expected to be equal to $(d-1)/d$ which
in two dimension gives $\alpha=1/2$ \cite{fiore2009,troster2012}.

\section{Conclusion}

We have analyzed the positivity of the heat capacity 
and emphasized this property as the condition for stability
of thermodynamic systems. We have shown that the slope
of the curves of energy versus temperature may not
coincide with the heat capacity. This is the case of
the calculations performed within the microcanonical
ensemble with periodic boundary conditions.
This point is understood if we
consider a microcanonical trajectory on the Gibbs
equilibrium surface. This surface of the thermodynamic
space spanned by the extensible variables has the convexity property.
However, the projection of a trajectory 
on the entropy-energy plane might lack convexity.
Analogously, the equation of state surface 
has the property of monotonicity but the projection
of a trajectory on the temperature-energy
plane might lack this property. The absence of monotonicity,
which is manifest by the negative slope is not in contradiction with the
positivity of the heat capacity because $\partial U/\partial T$ is not the
heat capacity.


\end{document}